\documentclass[aps,prl,superscriptaddress,amsfonts,amsmath,amssymb,reprint,showpacs,floatfix]{revtex4-1}

\usepackage{url}
\usepackage{bm}
\usepackage{graphicx}
\usepackage{amsmath}
\usepackage{amstext}
\usepackage{amssymb}
\usepackage{amsfonts}
\usepackage{amsbsy}
\usepackage{verbatim}
\usepackage{color}
\usepackage[colorlinks=true, urlcolor=blue, linkcolor=blue, citecolor=blue, pdftex]{hyperref}
\usepackage{multirow}
\usepackage{pdfpages}
\usepackage{floatrow}
\usepackage{gensymb}

\begin{document}

\title{Spin-$\frac{1}{2}$ Heisenberg $J_1$-$J_2$ antiferromagnet on the kagome lattice}

\author{\href{http://www.physik.uni-wuerzburg.de/institute_einrichtungen/institut_fuer_theoretische_physik_und_astrophysik/theoretische_physik_i/tp_i/mitarbeiter/dr_yasir_iqbal/}{Yasir Iqbal}}
\email[]{yiqbal@physik.uni-wuerzburg.de}
\affiliation{Institute for Theoretical Physics and Astrophysics, Julius-Maximilian's University of W\"urzburg, Am Hubland, D-97074 W\"urzburg, Germany}
\affiliation{The Abdus Salam International Centre for Theoretical Physics, P.O.~Box 586, I-34151 Trieste, Italy}
\author{\href{http://www.lpt.ups-tlse.fr/spip.php?article32}{Didier Poilblanc}}
\email[]{didier.poilblanc@irsamc.ups-tlse.fr}
\affiliation{Laboratoire de Physique Th\'eorique UMR-5152, CNRS and Universit\'e de Toulouse, F-31062 Toulouse, France}
\author{\href{http://people.sissa.it/~becca/}{Federico Becca}}
\email[]{becca@sissa.it}
\affiliation{Democritos National Simulation Center, Istituto Officina dei Materiali del CNR and 
SISSA-International School for Advanced Studies, Via Bonomea 265, I-34136 Trieste, Italy}

\date{\today}

\begin{abstract}
We report variational Monte Carlo calculations for the spin-$\frac{1}{2}$ Heisenberg model on the kagome 
lattice in the presence of both nearest-neighbor $J_1$ and next-nearest-neighbor $J_2$ 
antiferromagnetic superexchange couplings. Our approach is based upon Gutzwiller projected 
fermionic states that represent a flexible tool to describe quantum spin liquids with 
different properties (e.g., gapless and gapped). We show that, on finite clusters, a gapped 
$\mathbb{Z}_{2}$ spin liquid can be stabilized in the presence of a finite $J_2$ superexchange, 
with a substantial energy gain with respect to the gapless $U(1)$ Dirac spin liquid. However, 
this energy gain vanishes in the thermodynamic limit, implying that, at least within this 
approach, the $U(1)$ Dirac spin liquid remains stable in a relatively large region of the 
phase diagram. For $J_2/J_1 \gtrsim 0.3$, we find that a magnetically ordered state with 
${\bf q}={\bf 0}$ overcomes the magnetically disordered wave functions, suggesting the end 
of the putative gapless spin-liquid phase.
\end{abstract}


\maketitle

{\it Introduction}. In modern condensed matter physics, frustrated magnets provide us a window enabling a glimpse
of the vast and intriguing world of physics beyond the Landau symmetry-breaking and Fermi-liquid theories. One of the promising paths towards acquiring an understanding of this world 
is through the study of simple microscopic models. In this respect, the spin-$\frac{1}{2}$ Heisenberg 
antiferromagnetic model on the highly frustrated kagome lattice holds a distinguished position
by virtue of its promise in hosting a rich and exotic phase diagram, which is still attracting
substantial attention. However, a solution of this problem still proves to be an onerous task,
and indeed many studies in the past have emphasized the difficulty in reaching a final 
understanding of its ground-state and low-energy properties~\cite{Singh-1992,Lecheminant-1997,Misguich-2005,Sindzingre-2009,Nakano-2011,Lauchli-2011}.
A multitude of different ground states have been proposed, depending upon the approximate 
numerical and analytical techniques employed. A fully gapped $\mathbb{Z}_{2}$ topological 
spin-liquid ground state has been claimed for using density-matrix renormalization group 
(DMRG)~\cite{Yan-2011,Depenbrock-2012,Nishimoto-2013,Jiang-2013,Sheng-2014}, pseudofermion 
functional renormalization group~\cite{Suttner-2013}, and Schwinger boson mean-field 
calculations~\cite{Sachdev-1992,Wang-2006,Messio-2012,Punk-2014}. 
On the other hand, a gapless (algebraic) and fully symmetric $U(1)$ Dirac spin liquid has been
proposed as the ground state and widely studied using a variational Monte Carlo approach~\cite{Hastings-2000,Ran-2007,Hermele-2008,Ma-2008,Iqbal-2011a,Iqbal-2011b,Iqbal-2012,Iqbal-2013,Iqbal-2014,Bieri-2014}.
In addition, valence-bond crystals of different unit cell sizes and symmetries have been 
also suggested from other techniques~\cite{Marston-1991,Zeng-1995,Syro-2002,Nikolic-2003,Singh-2007,Poilblanc-2010,Evenbly-2010,Huh-2011,Hwang-2011,Poilblanc-2011,Capponi-2013}.
The coupled-cluster method suggested a ${\bf q}={\bf 0}$ (uniform) state~\cite{Gotze-2011}.
Finally, extending the construction of tensor network {\it Ans\"atze} of gapped $\mathbb{Z}_2$
spin liquids~\cite{Poilblanc-2012}, a recent calculation, based upon the so-called projected 
entangled simplex states that preserve lattice symmetries, gave remarkably accurate 
energies~\cite{Xie-2013}.

In this Rapid Communication, we focus on the spin-$\frac{1}{2}$ Heisenberg antiferromagnet in the presence of both 
nearest-neighbor ($J_1$) and next-nearest-neighbor ($J_2$) antiferromagnetic exchange 
couplings. Recent, state-of-the-art pseudofermion functional renormalization group 
studies~\cite{Suttner-2013} have claimed for a quantum paramagnetic ground state for 
$0 \leqslant J_2/J_1 \lesssim 0.7$, with the $J_2=0$ point corroborating a spin liquid with a 
small correlation length of about one lattice spacing, which is found to stay fairly similar 
as $J_2$ is turned on. This is followed by a region hosting a ${\bf q}={\bf 0}$ magnetically 
ordered phase, i.e., for $0.7 \lesssim J_2/J_1 \lesssim 1.5$. 
Finally, a nonmagnetic phase prevails again for $J_2 \gtrsim 1.5$, but its nature is unclear.
On the other hand, studies using projected Schwinger boson wave functions have suggested that,
at least on a small $36$-site cluster, the ${\bf q}={\bf 0}$ magnetically ordered state may be
defeated by a topological $\mathbb{Z}_{2}$ spin liquid for $J_2/J_1 \leqslant 1$~\cite{Tay-2011}.
Similar conclusions of a topological $\mathbb{Z}_2$ state are obtained for $J_2/J_1=0.1$ and 
$0.15$, by a measurement of the topological entanglement entropy using DMRG~\cite{Jiang-2013}.
More recent DMRG calculations pointed out that the transition from the quantum spin liquid
to the ${\bf q}={\bf 0}$ state may take place for relatively small values of 
$J_2/J_1$~\cite{Kolley-2014,Gong-2014}. In this regard, the issue of having magnetic order in 
the ground state of the $J_1$-$J_2$ model is still controversial and, so far, only few 
investigations have been done.

Here, we address the $J_1$-$J_2$ Heisenberg model within the realm of Gutzwiller projected 
Abrikosov fermion wave functions, by using state-of-the-art implementation of a variational 
Monte Carlo technique. In addition, we also consider the ${\bf q}={\bf 0}$ magnetic state, 
by using a Jastrow wave function, which represents an accurate way of describing ordered 
phases~\cite{Manousakis-1991}. For $J_2=0$, within the class of projected fermionic 
wave functions, there is strong evidence in support of a gapless scenario described by an 
algebraic $U(1)$ Dirac spin liquid. Indeed, explicit numerical calculations have shown the 
$U(1)$ Dirac spin liquid to be stable with respect to dimerizing into all known valence-bond 
crystal phases~\cite{Ran-2007,Ma-2008,Iqbal-2011a,Iqbal-2012}.
In addition, it was shown that, within this class of states, all the fully symmetric, gapped 
$\mathbb{Z}_{2}$ spin liquids have a higher energy compared to the $U(1)$ Dirac spin 
liquid~\cite{Iqbal-2011b,Lu-2011,Lu-2014,Tay-2011,Yang-2012}. Only a minor energy gain can be 
obtained by fully relaxing all the variational freedom of the wave function, namely, by a direct
optimization of the pairing function; however, this energy gain {\it decreases} upon increasing 
the cluster size~\cite{Clark-2013}. Most importantly, it was shown that upon application of a 
couple of Lanczos steps on the $U(1)$ Dirac spin liquid, very competitive energies can be 
achieved, still retaining a gapless state~\cite{Iqbal-2013,Iqbal-2014}. So far, a full 
treatment of the $J_1$-$J_2$ antiferromagnetic model has not been attempted within this 
approach.

%%%%%%%%%%%%%%%%%%%%%%%%%%%%%%%%%%%%%%%%%%%%%
\begin{figure}
\includegraphics[width=1.0\columnwidth]{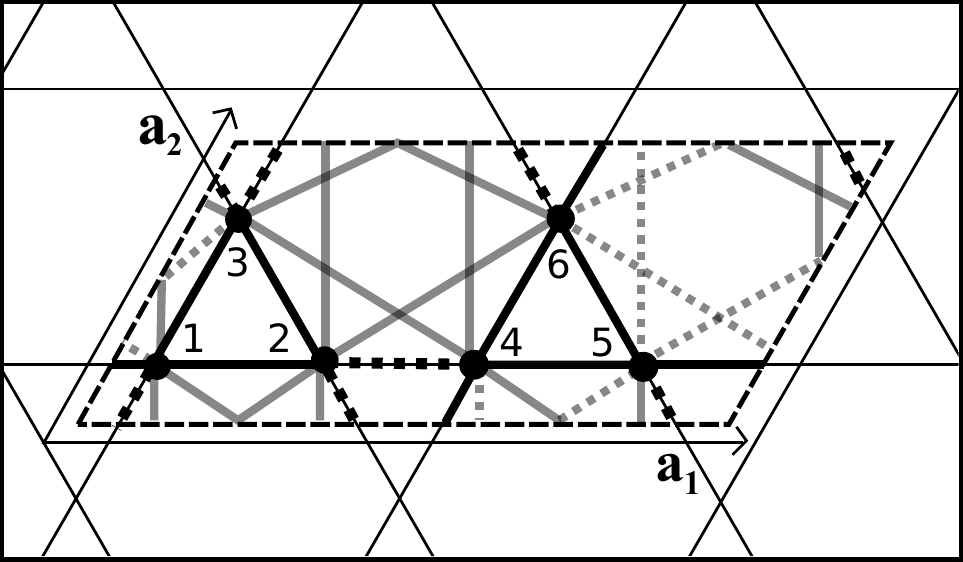}
\caption{\label{fig:kagome}
The $\mathbb{Z}_{2}[0,\pi]\beta$ spin-liquid {\it Ansatz}; black (gray) bonds denote 
nearest-neighbor real hopping (next-nearest-neighbor real hopping and real spinon pairing) 
terms; solid (dashed) black bonds have ${\rm s}_{ij}=1$ ($-1$) and solid (dashed) gray bonds 
have $\nu_{ij}=1$ ($-1$) [see Eq.~(\ref{eqn:MF-Z2})].}
\end{figure}
%%%%%%%%%%%%%%%%%%%%%%%%%%%%%%%%%%%%%%%%%%%%%

Here, we compute the variational energies of both the $S=0$ ground state and the first $S=2$ 
excitation, by considering spinon excitations around the Dirac nodes. We show that the best 
variational wave function is gapped for all the clusters that we can assess by our numerical 
technique. However, the energy difference between the gapped $\mathbb{Z}_2$ state and the
gapless $U(1)$ state decreases with increasing the size of the cluster and vanishes in the 
thermodynamic limit for all the values of $J_2$ that we have considered, i.e., 
$J_2/J_1 \leqslant 0.5$. Similarly, also the $S=2$ spin gap extrapolates to zero in the thermodynamic
limit. For $J_2/J_1 \gtrsim 0.3$, the magnetic Jastrow state overcomes the spin-liquid ones 
(both gapped and gapless), indicating that these kind of magnetically disordered states are 
no longer competitive. Here, we do not consider other quantum states, with topological 
or valence-bond order.

%%%%%%%%%%%%%%%%%%%%%%%%%%%%%%%%%%%%%%%%%%%%%
\begin{figure}
\includegraphics[width=1.0\columnwidth]{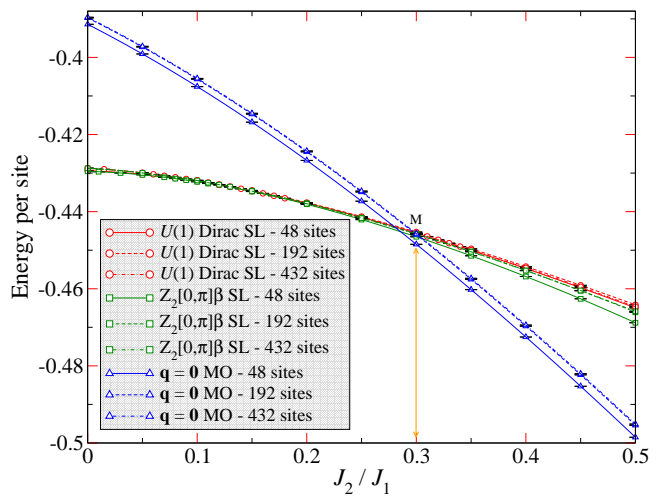}
\caption{\label{fig:comp}
(Color online) Energies per site as a function of $J_2/J_1$ for various competing phases are
shown on different cluster sizes. The point {\bf M} at $J_2/J_1=0.3$ marks the level crossing 
between spin-liquid (SL) and ${\bf q}={\bf 0}$ magnetically ordered (MO) phases.}
\end{figure}
%%%%%%%%%%%%%%%%%%%%%%%%%%%%%%%%%%%%%%%%%%%%%

{\it Model and Method}. The Hamiltonian for the spin-$\frac{1}{2}$ Heisenberg $J_1$-$J_2$ antiferromagnetic model is
\begin{equation}
\label{eqn:heis-ham}
{\cal H} = J_1 \sum_{\langle ij \rangle} \mathbf{S}_{i} \cdot \mathbf{S}_{j}
+J_2 \sum_{\boldsymbol\langle\langle ij \rangle\boldsymbol\rangle} \mathbf{S}_{i} \cdot \mathbf{S}_{j},
\end{equation} 
where both $J_1$ and $J_2>0$; $\langle ij \rangle$ and 
$\boldsymbol\langle\langle ij\rangle\boldsymbol\rangle$ denote sums over nearest-neighbor and 
next-nearest-neighbor pairs of sites, respectively. The $\mathbf{S}_{i}$ are spin-$\frac{1}{2}$ 
operators at each site $i$. All energies will be given in units of $J_1$.

%%%%%%%%%%%%%%%%%%%%%%%%%%%%%%%%%%%%%%%%%%%%%
\begin{figure*}[t]
\includegraphics[width=1.0\columnwidth]{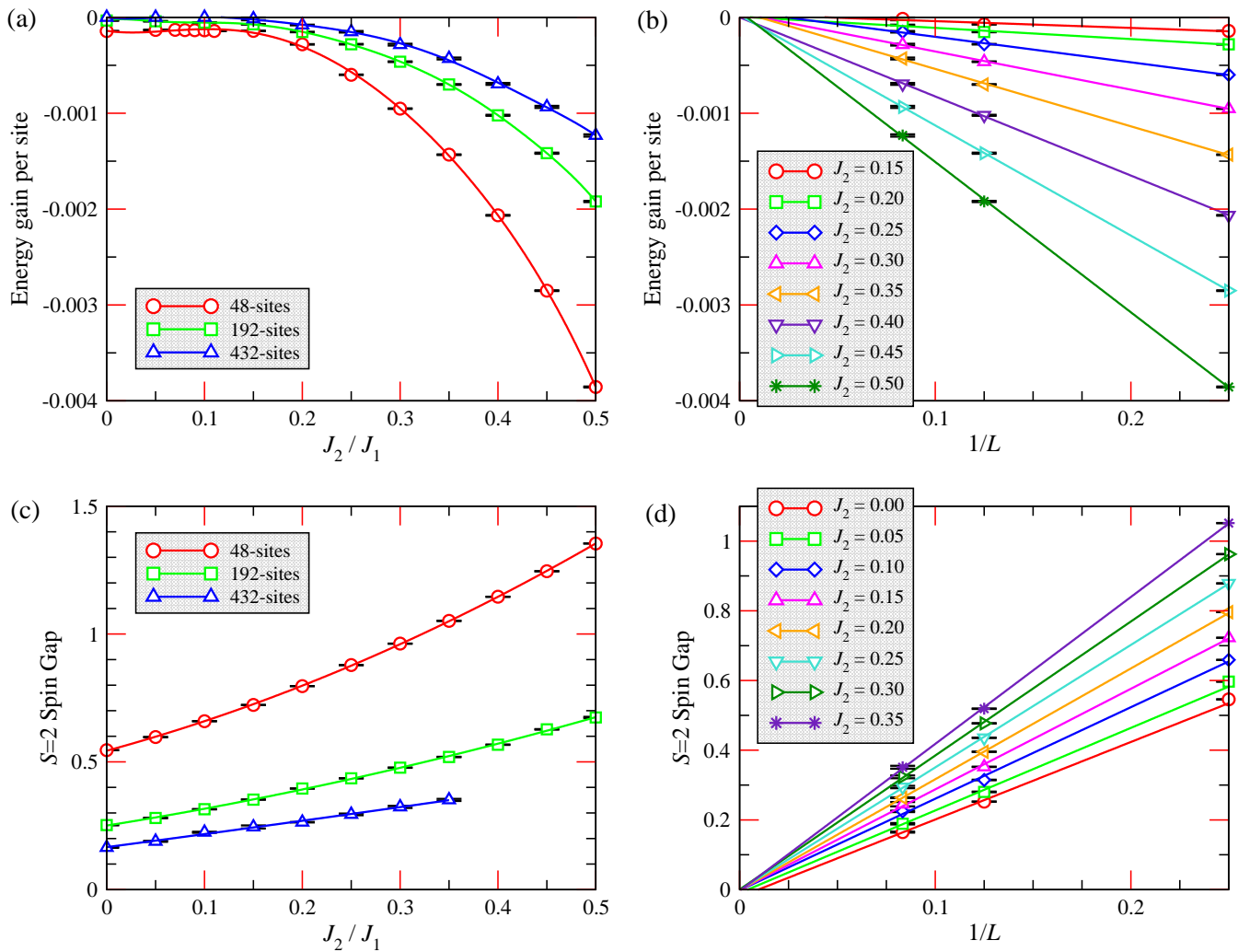}
\caption{\label{fig:fullres}
(Color online) (a) The gain in the energy per site of the $\mathbb{Z}_{2}[0,\pi]\beta$ 
state relative to the $U(1)$ Dirac spin liquid, for different cluster sizes, as a function 
of $J_2/J_1$ is shown. On the 48-, and 192-site clusters the gain remains finite down to 
$J_2=0$, whereas for the 432-site cluster it is zero (within error bars) for $J_2/J_1<0.15$. 
(b) The corresponding (linear) finite-size scaling of the energy gain per site is shown.
(c) The $S=2$ spin gap of the $\mathbb{Z}_{2}[0,\pi]\beta$ {\it Ansatz} for different clusters, 
as a function of $J_2/J_1$ is shown. (d) The corresponding (linear) finite-size scaling of the
$S=2$ spin gap is shown.}
\end{figure*}

\begin{figure}
\includegraphics[width=1.0\columnwidth]{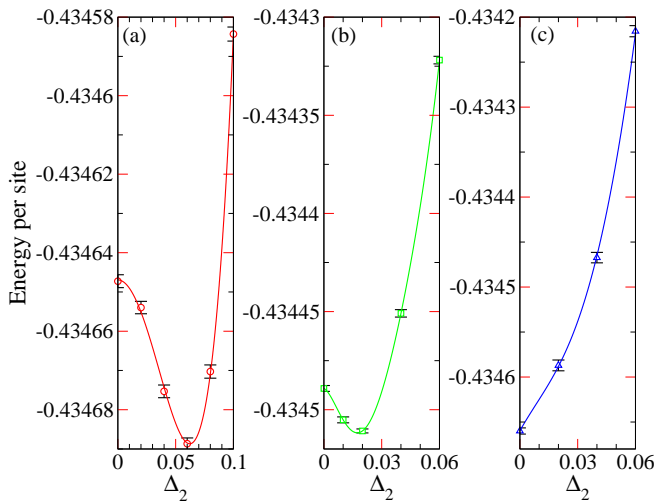}
\caption{\label{fig:landau}
(Color online) Energy per site of the $\mathbb{Z}_{2}[0,\pi]\beta$ state as a function of the
fermionic pairing $\Delta_2$, which leads to the lowering of the gauge structure from $U(1)$
to $\mathbb{Z}_{2}$ for $J_2/J_1=0.15$ and $L=4$ [left panel (a)], $L=8$ [middle panel (b)], 
and $L=12$ [right panel (c)].}
\end{figure}
%%%%%%%%%%%%%%%%%%%%%%%%%%%%%%%%%%%%%%%%%%%%

The variational wave function is constructed by considering a mean-field Hamiltonian that 
contains hopping and pairing. In particular, here we will focus on the so-called
$\mathbb{Z}_{2}[0,\pi]\beta$ state, as defined in Ref.~[\onlinecite{Lu-2011}]:
\begin{eqnarray}
&&{\cal H}_{{\rm MF}}\{\mathbb{Z}_{2}[0,\pi]\beta\} =
\chi_{1}\sum_{\langle ij\rangle,\alpha}{\rm s}_{ij} c_{i,\alpha}^{\dagger}c_{j,\alpha} \nonumber \\
&+& \sum_{\boldsymbol\langle\langle ij\rangle\boldsymbol\rangle}\nu_{ij}\Bigg\{\chi_{2}\sum_{\alpha}c^{\dagger}_{i,\alpha}c_{j,\alpha}+
\Delta_{2}(c^{\dagger}_{i,\uparrow}c^{\dagger}_{j,\downarrow}+ {\rm H.c.})\Bigg\}\nonumber \\
&+&\sum_{i}\Bigg\{\mu\sum_{\alpha}c_{i,\alpha}^{\dagger}c_{i,\alpha}
+\zeta_{\rm R} (c_{i,\uparrow}^{\dagger}c_{i,\downarrow}^{\dagger}+{\rm H.c.})\Bigg\},
\label{eqn:MF-Z2}
\end{eqnarray}
where ${\rm s}_{ij}$ and $\nu_{ij}$ encode the sign structure of the first and second 
nearest-neighbor bonds, respectively, as shown in Fig.~\ref{fig:kagome}. 
$c^{\dagger}_{i,\alpha}$ ($c_{i,\alpha}$) are the creation (annihilation) fermionic spinon 
operators at site $i$ with spin index $\alpha=\uparrow$,$\downarrow$. The real 
nearest-neighbor hopping ($\chi_{1}$) will be taken as a reference, and hence set to unity 
hereafter. This {\it Ansatz} is particularly interesting since it represents the only way 
of opening a gap in the $U(1)$ Dirac state, without breaking lattice symmetries. Indeed,
whenever pairing terms ($\Delta_{2}$ and $\zeta_{\rm R}$) vanish, the mean-field Hamiltonian 
reduces to the one defining the gapless $U(1)$ spin liquid. Given the extreme accuracy of the
latter state, the $\mathbb{Z}_{2}[0,\pi]\beta$ {\it Ansatz} has been considered for 
describing the topological liquid obtained by DMRG~\cite{Yan-2011,Depenbrock-2012}.

Other possible (gapless) $\mathbb{Z}_{2}$ {\it Ans\"atze}, suggested by the classification of 
Ref.~[\onlinecite{Lu-2011}], like the so-called $\mathbb{Z}_{2}[0,\pi]\alpha$ state, have also 
been studied by us, but they do not present any significant improvement with respect to the 
$U(1)$ {\it Ansatz}.

When a particle-hole transformation is performed on down electrons: 
\begin{eqnarray}
c^{\dagger}_{i,\downarrow} \to c_{i,\downarrow}, \\
c^{\dagger}_{i,\uparrow} \to c^{\dagger}_{i,\uparrow},
\end{eqnarray}
the mean-field Hamiltonian~(\ref{eqn:MF-Z2}) commutes with the total number of particles (while
it does not conserve the total spin along the $z$ axis). Therefore, the noncorrelated state is 
defined by filling suitable single-particle orbitals. Boundary conditions should be taken in 
order to have a unique state (i.e., filling all orbitals in a shell with the same mean-field 
energy). Here, we consider states with $S=0$ and $S=2$, both having ${\bf k}={\bf 0}$; these 
are particularly simple to handle, since they correspond to a single ``Slater'' determinant 
constructed by filling the lowest single-particle orbitals, i.e., 
$|\Psi_{\rm MF}(\chi_{2},\Delta_{2},\mu,\zeta_{R})\rangle$.

We would like to stress that the particle-hole transformation does not change the physical 
content of the model. Indeed, after this canonical transformation, the local Hilbert space of 
the spin model is changed into empty and doubly occupied sites (i.e., 
\textbar$\downarrow\rangle \to $ \textbar$0\rangle$ and \textbar$\uparrow\rangle \to$ \textbar$\uparrow \downarrow\rangle$), 
but the corresponding matrix elements of any operator are the same as in the original 
representation.

Then, in order to have a {\it bona fide} variational state for the spin model, the Gutzwiller 
projection $\mathcal{P}_{G}=\prod_{i}(1-n_{i,\uparrow}n_{i,\downarrow})$ must be applied,
enforcing the one fermion per site constraint, to the uncorrelated state:
\begin{equation}
\label{eqn:var-wf}
\text{\textbar}\Psi_{{\rm \mathbb{Z}_{2}[0,\pi]\beta}}(\chi_{2},\Delta_{2},\mu,\zeta_{R})\rangle=
\mathcal{P}_{G}\text{\textbar}\Psi_{\rm MF}(\chi_{2},\Delta_{2},\mu,\zeta_{R})\rangle.
\end{equation}

On the contrary, a simple and accurate variational wave function to describe magnetically 
ordered phases can be defined in terms of the original spins as~\cite{Manousakis-1991}
\begin{equation}
\label{eqn:var-cl}
\text{\textbar}\Psi_{{\rm Magnetic}}\rangle= \mathcal{J}_z \mathcal{P}_{S^z_{\rm tot}=0} \text{\textbar}{\rm SW}\rangle,
\end{equation}
where \textbar${\rm SW}\rangle$ is a spin wave state, described by a wave vector 
${\bf q}$ and a phase shift $\eta$ (one for each site in the unit cell):
\begin{equation}
\text{\textbar}{\rm SW}\rangle = 
\prod_{i} \left(|\downarrow\rangle_i + e^{\imath ({\bf q} \cdot {\bf R}_i + \eta_i)}
|\uparrow\rangle_i \right). \nonumber 
\end{equation}
$|{\rm SW}\rangle$ is equivalent to a classical state where each spin points in a given 
direction in the $XY$ plane. $\mathcal{P}_{S^z_{\rm tot}=0}$ is the projector onto the subspace
with $S^z=0$. Quantum fluctuations are included through the long-range Jastrow factor:
\begin{equation}
\mathcal{J}_z= \exp\left(\frac{1}{2} \sum_{ij} u_{ij}S^z_i S^z_j \right),
\end{equation}
where, in a translationally invariant system, the pseudopotential $u_{ij}$ depends on the 
distance $|{\bf R}_i-{\bf R}_j|$ of two sites. Here, we consider the case with 
${\bf q}={\bf 0}$, the three spins in the unit cell forming $120\degree$ with each other. 
All the independent parameters in the pseudopotential are optimized via Monte Carlo 
simulations.

{\it Results}. Our variational calculations are performed on square clusters (i.e., $3\times L \times L$)
with periodic boundaries in the spin Hamiltonian of Eq.~(\ref{eqn:heis-ham}). Let us start by a
comparison between spin-liquid and magnetic states. In Fig.~\ref{fig:comp}, we show the energy
per site for different cluster sizes (i.e., $L=4$, $8$, and $12$) for both the best gapped 
spin liquid and the gapless one, as well as for the optimized magnetic state with 
${\bf q}={\bf 0}$. In the presence of the next-nearest-neighbor coupling $J_2$, on small systems 
there is a finite energy gain in stabilizing spinon pairing. Moreover, this energy gain 
increases monotonically with $J_2$. The simple magnetic state, which is clearly unfavorable 
for small $J_2$, overcomes these spin-liquid states for $J_2/J_1 \gtrsim 0.3$. Whether a 
different kind of spin-liquid state or a valence-bond crystal may in turn overcome this 
magnetic state or not is an important problem that, however, we do not discuss here. We only 
want to mention that, unfortunately, the topological state proposed in Ref.~[\onlinecite{Tay-2011}] 
cannot be considered on large sizes, since a computation of permanents is required, so it is 
impossible to accurately estimate size effects. However, we would like to mention that
recent DMRG calculations~\cite{Kolley-2014,Gong-2014} pointed out that a magnetic state
with ${\bf q}={\bf 0}$ is obtained for small values of $J_2/J_1$, in rather good agreement
with our variational calculations.

In the following we restrict ourself to the region with small $J_2/J_1$, e.g., mainly
$J_2/J_1<0.3$, but also some slightly larger values, inside the putative magnetic region.
The actual energy gain due to spinon pairing is reported in Fig.~\ref{fig:fullres}(a) for
three different values of $L$. We obtain that on small systems (i.e., $48$- and $192$-site
clusters) a small energy gain is obtained for all values of $J_2$, down to $J_2=0$. Instead, 
for larger sizes, a finite next-nearest-neighbor coupling is needed to obtain a nonvanishing
energy gain due to spinon pairing. For $432$ sites, the best variational wave function is 
given by the $U(1)$ Dirac state for $J_2/J_1 \leqslant 0.15$, as previously reported by 
us~\cite{Iqbal-2011b}. By contrast, a sizable gain is obtained for larger values of $J_2/J_1$. 
The critical value of $J_2/J_1$, from which a nonzero energy gain is obtained, increases 
with increasing cluster size: for $768$ and $1200$ sites we obtain $J_2/J_1 \simeq 0.18$
and $0.20$, respectively. However, the size scaling of this quantity clearly indicates that 
the gain vanishes in the thermodynamic limit, for {\it all} the values of the 
next-nearest-neighbor superexchange coupling considered here [see Fig.~\ref{fig:fullres}(b)]. 
By considering both $S=0$ and $S=2$ variational states, we can assess the spin gap. Also for 
this quantity we obtain similar results [see Figs.~\ref{fig:fullres}(c) and~\ref{fig:fullres}(d)]. On finite 
clusters, the gap is finite and increases with $J_2/J_1$ but goes to zero when $L \to \infty$,
for {\it all} values of $J_2/J_1$ considered here. 

In order to better clarify the important changes of the energy landscape as a function of the 
cluster size, we report in Fig.~\ref{fig:landau} the variational energy of the 
$\mathbb{Z}_{2}[0,\pi]\beta$ state for different values of the pairing strength $\Delta_2$ 
(all the other variational parameters being optimized for the fixed value of $\Delta_2$) for 
$L=4$, $8$, and $12$ and $J_2/J_1=0.15$. The trend is clear: Both the optimal value of 
$\Delta_2$ and the energy gain with respect to $\Delta_2=0$ get systematically reduced, and 
eventually, for large enough size of the cluster, the minimal energy is obtained for the 
gapless $U(1)$ Dirac state (with $\Delta_2=0$).

These results shows that, at least within the Abrikosov fermion approach, the gapless 
$U(1)$ Dirac state is remarkably stable, not only for a particular point of the Heisenberg model 
(i.e., $J_2=0$) but in an entire region of the phase diagram. Moreover, our results suggest that 
the possible stabilization of a $\mathbb{Z}_{2}$ topological spin liquid found by DMRG calculations 
in the presence of a small $J_2/J_1$~\cite{Jiang-2013} may possibly be due to the finiteness of the 
cluster.

{\it Conclusions}. In summary, we have shown that the gapless spin-liquid state, described by the $U(1)$ 
{\it Ansatz} of Ref.~[\onlinecite{Ran-2007}], is remarkably stable also when a next-nearest-neighbor 
antiferromagnetic coupling $J_2$ is considered in the Heisenberg model. Interestingly, on finite 
clusters, a notable energy gain may be obtained by allowing spinon pairing that opens a gap in the 
mean-field spectrum and lowers the gauge structure down to $\mathbb{Z}_{2}$; however, we identify 
this to be an artifact due to finite-size effects, since the energy gain vanishes in the thermodynamic 
limit, giving back a state with Dirac nodes and, therefore, gapless excitations. It is worth 
mentioning that very recent state-of-the-art DMRG calculations also provide an inkling of a gapless 
ground state~\cite{Zhu-2014}.

Finally, we would like to remark that, performing the Lanczos steps procedure that has been 
used previously~\cite{Iqbal-2013,Iqbal-2014}, we have evidence that the $U(1)$ Dirac state 
gives a perfectly stable and {\it linear} convergence to the ground state upon performing a 
zero-variance extrapolation. This implies a large overlap and a close connection of the $U(1)$
Dirac state to the true ground state, similar to what has been obtained for $J_2=0$. On the 
contrary, upon starting from the gapped $\mathbb{Z}_{2}[0,\pi]\beta$ wave function, we have, 
for both $S=0$ and $S=2$ states, large statistical fluctuations and consequently a large 
variance of energy, especially at the second Lanczos step level. Most importantly, preliminary
calculations show that the variance of energy either remains constant (for $S=0$) or even increases 
(for $S=2$) at the second Lanczos step compared to the first Lanczos step~\cite{Becca-2014}.
These facts may indicate that this gapped wave function may have a considerable overlap with 
excited states, thus implying that it is not a faithful representation of the true ground state.

{\it Acknowledgments}. We thank W.-J. Hu, S.-S. Gong, and D. N. Sheng for interesting discussions. This research was 
supported in part by PRIN 2010-11. D.P. acknowledges support from NQPTP Grant No. ANR-0406-01 from the 
French Research Council (ANR). Y.I. acknowledges Stefan Depenbrock for stimulating
discussions during the NORDITA program ``Novel Directions in Frustrated and Critical 
Magnetism'' (2014).

\newpage

\end{document}